\documentclass[twocolumn,showpacs,preprintnumbers,amsmath,amssymb]{revtex4}
\usepackage{graphicx}
\usepackage{dcolumn}
\usepackage{bm}
\usepackage{soul,color}
\begin{document}
\title{Modelling Financial Markets
by Self-Organized Criticality}

\author{Alessio Emanuele Biondo $^1$, Alessandro Pluchino $^2$, Andrea Rapisarda$^2$}
\bigskip 

\affiliation{$^1$ Dipartimento di Economia e Impresa - Universit\'a di Catania, Corso Italia 55, 95129 Catania, Italy\\
$^2$ Dipartimento di Fisica e Astronomia, 
Universit\'a di Catania  and INFN sezione di Catania, 
Via S. Sofia 64, 95123 Catania, Italy}
\date{\today}
\bigskip
\begin{abstract}
\noindent We present a financial market model, characterized by self-organized criticality, that is able to generate endogenously a realistic price dynamics and  to reproduce well-known stylized facts. We consider a community of heterogeneous traders, composed by chartists and fundamentalists, and focus on the role of informative pressure on market participants, showing how the spreading of information, based on a realistic imitative behavior, drives contagion and causes market fragility. In this model imitation is not intended as a change in the agent's group of origin, but is referred only to the price formation process. We introduce in the community also a variable number of random traders in order to study their possible beneficial role in stabilizing the market, as found in other studies. Finally we also suggest some counterintuitive policy strategies able to dampen fluctuations by means of a partial reduction of information.
\end{abstract}

\pacs{89.65.Gh,89.65.Gh, 05.65.+b 
}
\maketitle
\vspace{0.25cm}

\section{Introduction}

 Price dynamics in financial markets is the result of the interactions 
 and mutual feedbacks of many interconnected  agents, who trade according to their information. In general,  it shows a very complex and hardly predictable behavior, which  is not easy  to simulate and control. 

Financial integration on a global scale is today so extreme that policy-makers need to learn how to prevent dangerous dynamics. In macroeconomic terms, the current ``mainstream" economics approach has shown to be ineffective in taming the wild fluctuations that financial markets often show. A very simple example of how urgent new policy designs are, is given by the evidence of scarce effectiveness of the well-known ``Tobin Tax" on financial transactions \cite{cme-group-report}. 
In other words, the neoclassical economic approach seems to have failed its mission at the macroeconomic level: neither the idea of efficient markets based on perfect equilibrium \cite{fama1970efficient}, nor the rational expectations paradigm \cite{lucas1972expectations, sargent1976rational}, have helped understanding the aggregate financial behavior and the reasons at the core of severe financial crises.

Recently, many studies presented intriguing insights on the characterization of social systems as complex entities, suggesting the fruitful adoption of tools and tecniques coming from statistical and theoretical physics\cite{mantegna1999introduction, Helbing1,  Helbing}. 
New promising directions of research to model financial markets are based on the concepts of bounded rationality and behavioral heterogeneity, i.e. agents can have a limited rationality and not all of them assume the same behavior. In addition,  the topological network structure that characterizes economic interactions at the macroeconomic level has been discovered to have a crucial role.   This kind of approaches focus also on the role of information and of contagion spreading. From this perspective, individual choices within a social context can reveal to be driven more by rules of thumb than by perfect knowledge and optimal computational ability \cite{simon1957models,tversky1974judgment}. Human interactions and individual psychology of traders cannot be ignored any longer, as  dramatically shown in many situations \cite{akerlof2010animal}.
A more realistic description of  financial markets  presumes that agents are not fully rational in the sense described by the mainstream of rational expectations literature. The many problems of the rigid approach of full perfect rationality are discussed in a number of papers \cite{camerer2003,barberis2003survey,colander2009financial,kahneman1979prospect}. 

In a series of recent studies we have taken into account this limited rational behavior of agents operating in a complex network structure and have already shown evidence that these alternative approaches may reveal  useful applications. As an example, the beneficial role of random strategies has been  described in several papers for socio-economic systems \cite{AAPeter1, AAPeter2, AAParl}, and in particular for financial markets \cite{noiJSP, noiPRE, noiPLOSONE, noiCP}. A simulative agent-based methodology will hopefully lead to more advances in understanding complex economic dynamics  and in policy design \cite{gatti2011macroeconomics}.

A very wide literature deals with the agent-based approach for financial markets simulations, \cite{brock1997models,brock1997rational,brock1998heterogeneous,chiarella1992dynamics,chiarella2001asset,day1990bulls,franke1998cautious,hommes2001financial,lux1995herd,lux1998socio,lux1999scaling}. In particular, the Heterogenous Agent Models (\textit{HAM}) represent a fruitful approach able to study the complex interactions of \emph{different} individuals with \emph{different} behaviors. In the greatest part of the HAM-related literature, surveyed in refs.  \cite{lebaron2006agent} and \cite{lux2009economics}, agents are divided in two typical categories: \textit{fundamentalists} and \textit{chartists}. Fundamentalists are traders with an eye on the fundamental value of assets; for example, they form their opinions and their strategies to decide whether to buy a share or not, by looking at its current price level and by comparing it with its fundamental values (that is, roughly speaking, almost always the present discounted value of future expected dividends). On the contrary, chartists are technical analysts, who form their expectations on assets prices and decide their strategies by looking at the charts, i.e. at trends and graphic dynamics of past prices.

The existing literature has usually described the imitation on financial markets by assuming that a trader can switch group, from fundamentalists to chartists or vice versa \cite{cristelli2014}. We propose, instead, to refer to the trading decision of the trader, i.e. to the price: the agent who decides to imitate a trader, simply follows the price prevision assumed by that trader, no matter which group the latter belongs to. Thus, the persuasive strength of information may induce, say, a chartist to imitate the price set by a fundamentalist without switching group. As a result of this diffusion of informative signals, according to the topology of the network, extreme phenomena may spontaneously emerge. 

Differently from other attempts to describe herding in financial markets \cite{alfaranoluxwagner,kononoviciusgontis13}, our model considers the pressure coming from the accumulation of information, by recalling some features of a model of earthquakes \cite{olami} and presents a number of key-features: (i) an endogenous price setting mechanism that can reproduce all relevant stylized facts of true financial markets, (ii) heterogenous agents organized in different groups, with a realistic imitative behavior, (iii) an emergent aggregate dynamics that suitably describes extreme events involving market participants as in true financial bubbles/crashes. In the following we present the model in detail and discuss its properties. 

The paper is organized as  follows: in section 2 the model is described and the reproduction of stylized facts is addressed, also in comparison with a real dataset; in section 3 the role of random traders on financial crises and some policy suggestions are discussed; finally, in section 4  conclusions are drawn. 

\section{The Model}

Financial markets are populated by interacting agents, who continuously look for new information and try to update their expectation models trying to obtain accurate forecasts. However, the system exhibits unavoidably complex characteristics, since individual beliefs and decisions depend on those of others. Thus, facing unpredictability, agents are forced to act inductively, by using different and volatile strategies, continuously updated according to credibility of signals that they receive from the market itself. In previous studies, links between extreme events in financial markets and the dynamics of informative cascades have been investigated, \cite{noiPRE, noiCP}. However, at variance with previous models, the present model considers a realistic feedback mechanism that let the heterogenous traders determine prices under the influence of prices dynamics itself. For these reasons, we call it \emph{Contagion Financial Pricing} (CFP henceforth) model.

\subsection{Setting Description}

The CFP model here presented describes  an artificial financial market with  a population of $N$ investors. These agents $A_i$ (with $i=1,...,N$) are connected among themselves in a \emph{Small World} (SW) network, usually adopted to describe realistic communities in social or economical contexts \cite{watts1999}. The SW network here considered, shown in Fig.1, is obtained from a square $2$-dimensional regular lattice, with open boundary conditions, by randomly rewiring its short-range links with a probability $p=0.02$, therefore creating a given number of long-range links. The final average degree of the network is equal to $<k>=4$. See ref. \cite{noiPRE} for more details. 

\begin{figure}[t]
\includegraphics[width=3.0in,angle=0]{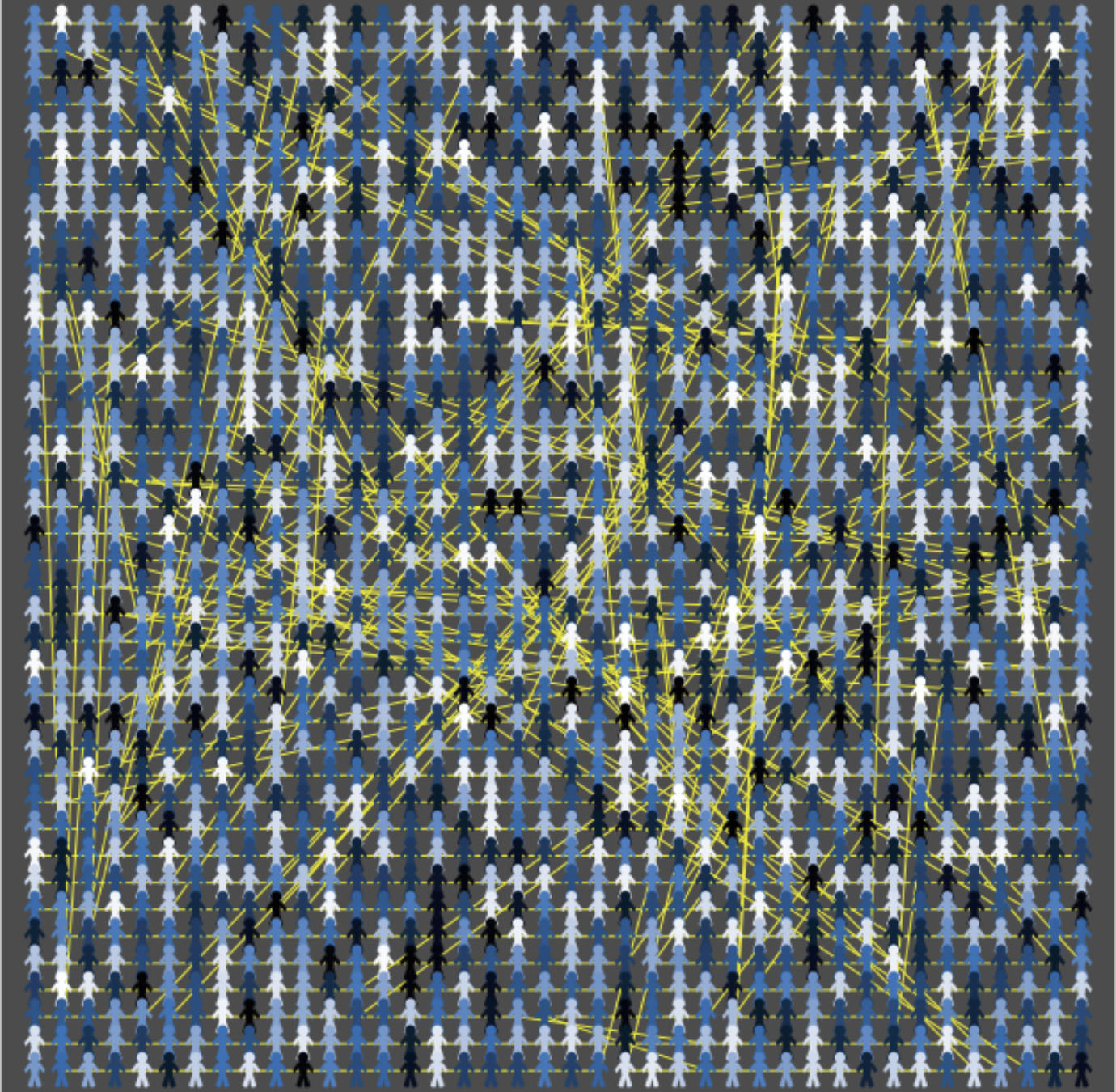} 
\caption{(Color online) An example of the 2D Small World lattice adopted in our model (with $n=40$). Traders are distributed on a square network where short- and long-distance links are visible. Agents are coloured differently in order to represent their levels of information: the brighter a trader is, the more informed she is. Initial levels of information are distributed randomly. See text for further details.}
\end{figure}

Since we do not consider any portfolio, the model implicitly assume that an ideal counter-part always exists at each time-step. Our main goal is to study  the role of  composition of the population of traders and that of the spreading of information in influencing the market dynamics, its stability and the probability of bubbles and crashes. Because of this reason, the price time series that our model generates, $p_{t}$ where $t$ is the time, has to be considered as the result of the transactions occurring among the traders, even if we do not describe either the order-book or the portfolio dynamics of investors. The population is characterized by the existence of three groups of traders: \emph{(i)} fundamentalists, \emph{(ii)} chartists, and \emph{(iii)} random traders. 

More precisely, a fundamentalist bears in mind a \emph{fundamental value} that she believes is the ``correct" value of the asset being traded: she believes that the market dynamics will tend to let this correct value prevail. Thus, she participates to transactions by stating her price $p^{f}_{t+1}$ on the basis of the discrepancy between the last observed price, $p_t$, and this fundamental value, $p_f$: 
\begin{equation}
p^{f}_{t+1}=p_t+\phi (p_f - p_t) + \epsilon
\end{equation}
In such a way, the fundamentalist's individual price will be greater or smaller than the previous market price according to the fact that the fundamental value is greater or smaller than the previous market price itself. The parameter $\phi$ is a sensitivity parameter that regulates how much of the discrepancy will be embedded in the new price. Finally, $\epsilon$ is a stochastic noise term, randomly chosen in the interval $(-\sigma, \sigma)$, with $\sigma$ fixed at the beginning of simulations and extraction done with uniform probability. It is worth to notice that the value of $\phi$ can either be fixed or, in order to gain heterogeneous behavior, different for any fundamentalist. In this case, it will be normally distributed with given mean and standard-deviation.

A chartist, instead, is a technical analyst and decides her behavior according to her inspection of charts of past prices. Therefore, the next individual price that she will state on the market will be a function of past prices. The simplest (and less arbitrary) function that we have chosen is the average of last $M$ prices. Thus, a chartist will decide ``her" price according to the circumstance that the last market price is greater or smaller than the difference between the price and the average of last $M$-values, $p_M$. More precisely,
\begin{equation}
p^{c}_{t+1}=p_t+\frac{\kappa}{M} (p_t - p_M) + \epsilon
\end{equation}
Also in this case, the behavioral heterogeneity can be obtained by letting both $M$ and $\kappa$ (respectively the \emph{length} of trader's retrospective sight and the sensitivity of forecasts to past prices) to be extracted from a normal distribution with previously fixed mean and standard deviation values. Again, $\epsilon$ is a stochastic noise term defined as in Eq.(1). Chartists will also be considered with two different attitudes: trend-following or trend-reversal (such a specification is easily obtained by considering negative values for $\kappa$). 

Finally, we consider also a third category, that of random trading agents. A random trader is defined as an investor who does not care at all about previous or fundamental values and select her price, $p^{r}_{t+1}$, by choosing it randomly from a uniform distribution of values, ranging from $0$ and the last market price value, i.e.:
\begin{equation}
p^{r}_{t+1} \in [0, p_t]
\end{equation}

The global market price, $p_{t+1}$, will be obtained as the weighted average of individual prices, the weight being the proportion of each group relative to the total population
\begin{equation}
p_{t+1}=\frac{F}{N}\sum p^{f}_{t+1}+\frac{C}{N}\sum p^{c}_{t+1}+\frac{R}{N}\sum p^{r}_{t+1} + \omega
\end{equation}
where  $F$ is the total number of fundamentalists, $C$ is the total number of chartists, $R$ is the total number of random-traders, and $N=F+C+R$ is the total population of agents. Finally, $\omega$ is a global noise term that is related to the information accumulated by traders, as defined below.

Consistently with previous studies, we depict a situation where each agent is exposed to two streams of informative pressures, a {\it global} one ($a$) and an {\it individual} one ($b$) \cite{noiPRE, noiCP}: 
\\
\\
($a$) The first comes from the market and represents the general climate that each trader perceives from news about the current states of markets. We model this phenomenon by associating to each trader a real variable $I_i(t)$ $(i=1,2,...,N)$, that represents the information possessed at time $t$. At the beginning of each simulation (at $t=0$), the informative level of traders is set to a random value in the interval $(0,I_{th})$, where $I_{th}=1.0$ is a threshold value that is assumed to be the same for all agents. Then, the simulation starts and all traders receive a global informative pressure, which reaches them uniformly. In other words, each investor acts as a sort of accumulator of information: at each time-step $t > 0$, the information accumulated by all traders is increased by a quantity $\delta I_i$, different for each agent and randomly extracted within the interval $[0,(I_{th}-I_{max}(t))]$, where $I_{max}(t)= max \{I_i(t)\}$ is the maximum value of the agents' information at time $t$, and each trader sets her new price following equations (1), (2) or (3). Finally, the global market price $p_{t+1}$ follows from equation (4), where the global noise is assumed to be $\omega=\epsilon \ e^{\beta I_{av}(t)}$, where $\epsilon$ is the same noise term as in Eqs.(1) and (2), $\beta$ is a constant chosen in a suitable interval and $I_{av}(t)$ is the average value of the information accumulated by all the traders at time $t$. Thus, the global price formation is affected in a non-linear (and stochastic) way by the total information present in the system.
\\
\\
($b$) The second one, is an individual transmission that every trader receives from her close neighbors (i.e. from the other known traders). Actually, when a given agent $A_k$ accumulates, from the general flow ($a$) of the market, enough information to exceed her personal threshold value $I_{th}$, she becomes ``active''. At this point it is important to distinguish non random traders from random ones:

\begin{itemize}  
\item when a given non random trader $A_k$ (fundamentalist or chartist) surpasses her threshold at time $t$, immediately after fixing her new price $p^k_{t+1}$ she transmits the informative signal to her neighbors within the trading network according to the following simple herding mechanism, analogous to the energy transmission in earthquake dynamics, see \cite{noiPRE}:
\begin{equation}   
\label{av_dyn}       
I_k > I_{th}  \Rightarrow \left\{ 
	\begin{array}{l}
       I_k \rightarrow 0, \\
       I_{nn} \rightarrow I_{nn} + \frac{\alpha}{N_{nn}} I_k,
       \end{array} 
	\right.
\end{equation}
where ``nn'' denotes the set of nearest-neighbors of the active agent $A_k$. $N_{nn}$ is the number of direct neighbors, and the parameter $\alpha$ controls the level of dissipation of the information during the dynamics ($\alpha=1$ corresponds to the conservative case, but in our simulations we always adopted values strictly less than 1, in analogy with \cite{olami}).  
As a consequence of the received amount of information, someone of the involved neighbors may become active too and pass the threshold level as well, thus transmitting, in turn, her signal to her neighbors and so on: in this case we say that an informative avalanche started and we call this process, which can involve a variable number (even very high) of non random active agents, a ``financial avalanche". The central point is that all the agents involved in the financial avalanche will \emph{imitate} the price $p^k_{t+1}$ set by the former one, who originated the avalanche, regardless of their own group (fundamentalist or chartist); the reader should be aware that, in such a way, we do not consider any dynamical change in the population composition: instead, we want to focus on the more realistic definition of imitation, that keeps unchanged the ``type" of the trader even if let her copy the trading decision of one of her neighbors;

\item on the other hand, random traders are only affected by the general climate ($a$) of the market but do not  influence the other traders nor are influenced by them; 
we will show, as already discussed in previous studies within a different approach (\cite{noiPRE, noiCP}) that their role is crucial for a damping of the herding avalanches and reduce the volatility of the time series. 

\end{itemize} 

In the next paragraph we will show in  detail, through a first set of single-event numerical simulations, how such a dynamical model shows self-organized criticality (SOC) (see also \cite{caruso}), which influences  the global price formation.  We will also discuss  the extent to which our model is able to reproduce the so-called {\it stylized facts}, characteristic of real price time series. No random traders will be considered at the beginning.

\begin{figure}[t]
\centering 
\includegraphics[width=3.2in,angle=0]{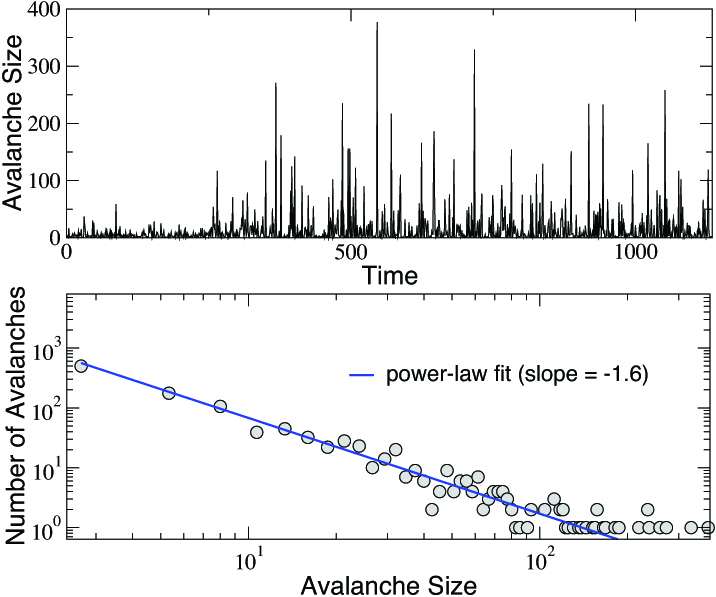}
\caption{(Color online) In the top panel, the time series of the avalanches generated by the model is reported whereas, in the bottom panel, the pdf  of the sizes of avalanches is shown. The latter can be fitted by a  power-law curve,  also reported, with slope $-1.6$. See text for further details.}
\end{figure}

\subsection{Reproduction of stylized facts and comparison with real data}

Multi-agent models of financial markets have to be able to replicate some specific features, known as \emph{stylized facts} \cite{kaldor1961capital}. In other words, before being used as a valid specimen lab, any simulative model  should manifest the ability to reproduce known characteristics of real financial markets. In order to check this within the context of the CFP model, we adopted the following settings:
\begin{itemize}
\item[--] fundamentalists base their fundamental price on a completely external, pre-set value;
\item[--] heterogeneity of traders has been specified by means of personal parameters that differentiate behavioral values of different traders, also within each group;
\item[--] different population composition and different network sizes are considered.
\end{itemize}

Let us first consider a community of $N=1600$ traders (no random traders are considered for the moment), connected as in Fig.1 and divided in $F=400$ fundamentalists and $C=1200$ chartists. At $t=0$, as previously explained, each trader starts with a given random amount of information $I_i(t)\in (0,1)$ and, at each time-step $t > 0$, receives a further (random) amount of global information $\delta I_i$. Then, all traders fix their price following equations (1) or (2), where the values of parameters are: $p_f=5000$ (fundamental price), $\phi=2.0$ (fundamentalists' sensitivity parameter), $\sigma=200$ (amplitude of stochastic noise), $M\in[0,90]$ (length of retrospective sight), $\kappa=2.0$ (chartists' sensitivity of forecasts to past prices). Finally, the next global market price $p_{t+1}$ is calculated by means of equation (4), with $\beta=16$ (exponent of the global noise term). 

In the following, we compare the stylized facts obtained for a CFP global price time series of $10000$ iterations, with the analogous ones obtained for a real time series of comparable length, in particular for the General Electric (GE) stock prices, collected day by day from $01/01/1962$ to $14/03/2014$ \cite{generalelectric}. During a single run of the CFP model, a given number of informative avalanches occur, following the herding mechanism (5) with $\alpha=0.92$. In Fig.2 we show the sizes of these avalanches versus time (top panel), together with their frequency distribution (bottom panel), whose clear power law behavior confirms the SOC-like character of the herding dynamics. Such an internal feature, combined with the global informative pressure expression of the market climate, strongly affects the emerging global price series shown in panel (b) of Fig.3: after an initial positive trend, corresponding to the initial transient dynamics which precedes the triggering of the critical state (see top panel of Fig.2), the global price values start to strongly fluctuate, as also proved by the corresponding fluctuating behavior of the returns time series (panel d). Recall that, given a time series $p_{t}$, returns $r_{t}$ are defined as: $r_{t}=log(p_{t+1})-log(p_{t})$. From the comparison with the GE stock prices and returns time series (panels a and c, respectively), we conclude that the simulation results closely mimics typical characters of real financial markets.                     

\begin{figure}[t]
\centering 
\includegraphics[width=3.4in,angle=0]{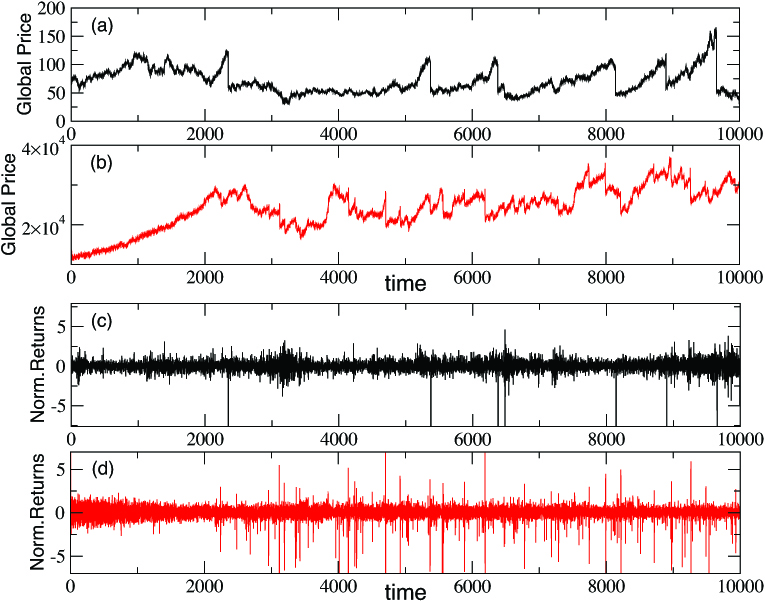} 
\caption{(Color online) The simulated global price time series, $p_t$ panel (b) and the correspondent returns time series $r_t$, panel (d), for the CFP model are compared with the historical time  series for the General Electric stock prices, panels (a,c). See text for further details.}
\end{figure}

The stylized facts that are usually reported and verified in true price series and that we successfully tested in our model, are \cite{chakraborti2009econophysics}: 
\\
1. Fat Tails of Distribution of Returns 1
\\
2. Absence of Auto-Correlations of Returns
\\
3. Volatility Clustering

\begin{figure}[t]
\centering 
\includegraphics[width=3.1in,angle=0]{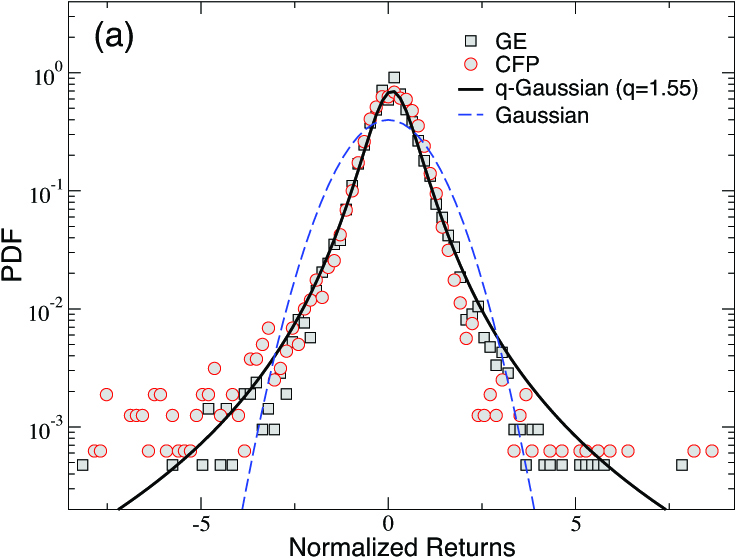}
\includegraphics[width=3.2in,angle=0]{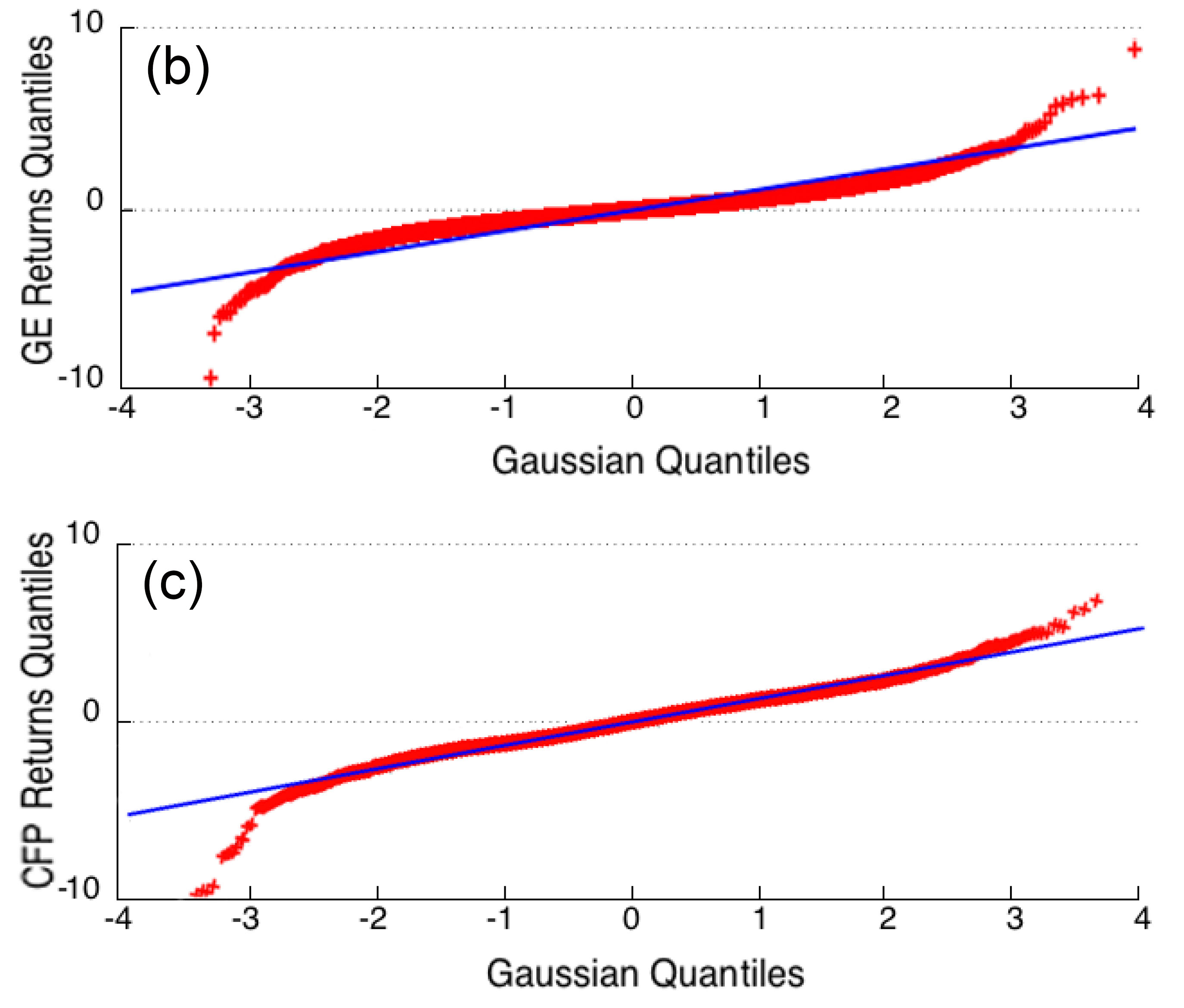} 
\caption{(Color online) Stylized Fact \#1. Panel $(a)$: PDFs of both GE (squares) and CFP normalized Returns in comparison with a Gaussian with unitary variance (dashed blue line). Evidence of non-Gaussian behavior emerges, due to the presence of fat tails that can be fitted by a q-Gaussian curve, see text. Panels $(b)$ and $(c)$:  the $QQ-plots$ of, respectively, GE and CFP Returns quantiles (cross shapes) compared with those of a Gaussian (straight line). Again strong deviations from normal behavior are visible.}
\end{figure}

\subsubsection{Fat Tails Distribution of Returns}

It is well known that financial returns distributions are non-Gaussian curves and, in particular, leptokurtic and asymmetric \cite{chakraborti2009econophysics}. In Fig.4(a) we show such a distribution for both a price series generated by our model (open circles) and for the GE stock price one (open squares). In particular, we consider here normalized returns, defined as \mbox{$(r_{t} - r_{av})/r_{stdev}$} (where $ r_{av}$ and $ r_{stdev}$ are, respectively, mean and standard deviation calculated over the whole returns series). 

\begin{figure}[t]
\centering 
\includegraphics[width=2.5in,angle=0]{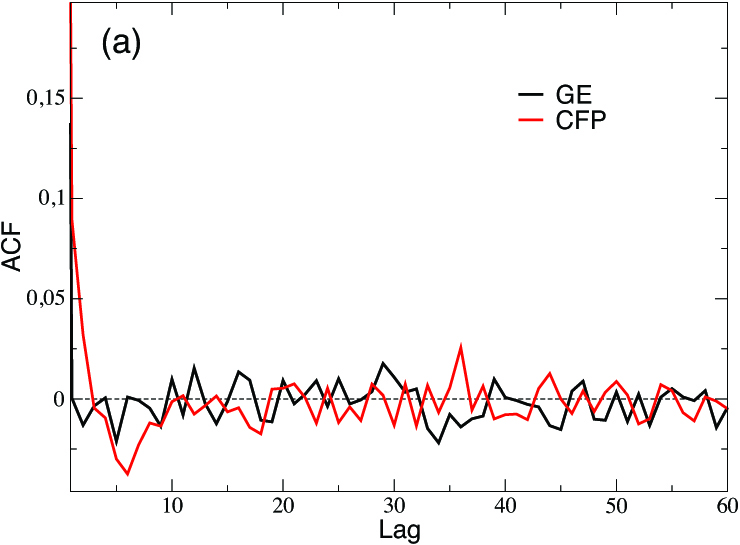}
\includegraphics[width=2.5in,angle=0]{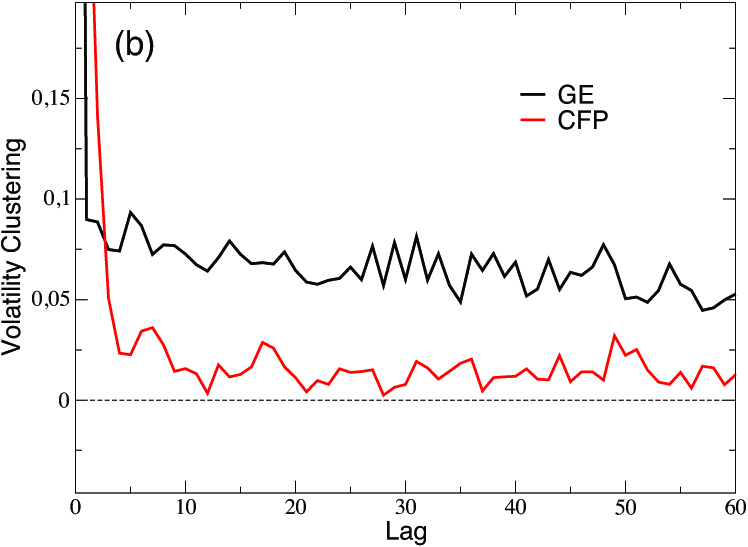}
\caption{(Color online) Stylized Fact \#2. Panel (a): Auto Correlation Function (ACF) of both GE and CFP Returns Series shows no significative autocorrelation of the returns. Stylized Fact \#3. Panel (b): ACF of both GE and CFP Absolute Returns Series show an autocorrelation of absolute returns which slowly decays towards zero remaining positive for all the lag intervals considered.}
\end{figure}

Simulation and real data are also compared with a standard Gaussian with unitary variance (dashed blue curve): fat tails and asymmetry are well visible in both the returns distributions, that can be fitted by a $q$-Gaussian, defined as $G_{q}=A {[1-(1-q) B x^2]}^{1/(1-q)} $. The $q$-Gaussian is a curve with power-law tails, defined in the context of non-extensive statistical mechanics which has been  widely used also in economics and in coupled systems at the edge-of-chaos\cite{tsallis1,tsallis2,tsallis3,tsallis4,miritello1}. The entropic index $q$ measures deviations from Gaussian behavior, for $q=1$ a standard Gaussian, with exponential tails, is recovered. In our  case, $q=1.55$  while the values of the other fitting parameters are $A=0.7, B=3$. This point seems  very interesting and deserves further investigation,  but it is beyond the scope of the present paper and it will be explored elsewhere. 

In Fig.4 (b) and (c) we present the QQ-plot of the returns obtained from both, respectively, the GE and CFP price series. This kind of graph compares quantiles of a distribution with quantiles of the Gaussian. The straight line $y=x$ is the test benchmark, since it represents the case of a distribution that behaves normally. The cross shapes curves in both the panels, clearly deviating from linearity, confirms the presence of fat tails and therefore the non Gaussian behavior of the returns distributions.

\subsubsection{Absence of Auto-Correlations of Returns and Volatility Clustering}

The absence of autocorrelations of returns is sometimes referred to as the absence of simple arbitrage possibilities: it is essentially equivalent to say that on average is not possible to foresee the price variation from $t$ to $t+1$. Thus, profits may derive just from risky investments (sometimes traders describe this occurrence by saying that there are no \textit{free lunches}). 

In Fig.5(a) we report the Auto Correlation Function (ACF) for both the GE and the CFP Returns Series. We observe that, as it has been widely documented in refs. \cite{pagan1996econometrics} and in \cite{cont1997scaling}, among others, for true returns series of financial assets, the ACF calculated from tick transactions prices of the CFP returns series shows very similar results to those obtained for the real GE stock returns, with no evidence of correlations. In this regard, it is statistically relevant to talk about the ``correct" observational timing of financial markets. In what we discuss, we will always consider \emph{transactions prices} in order to refer to trade time: each time a price is set (in other words, this represent the moment when a transaction is done in our model), time counter is increased by $1$. In our framework, where an order book is missing, this is also the closest approximation to \emph{tick prices} which are generated, nowadays, even in milliseconds.  The \emph{tick} is the smallest variation possible of a financial price: it represents the unit measure of the price variation. In the trading jargon, a dealer looks how many \emph{ticks} an asset has gained/lost. 

Finally, we observe that the absence of correlation among returns does not imply the stability of properties of the distribution with respect to time. In particular, one of the most important \emph{facts} is the circumstance that absolute returns exhibit a long-range slowly decaying autocorrelation function. This property has been named ``volatility clustering" and described in ref. \cite{mandelbrotvariation63}. In Fig.5(b) we plot the ACF of absolute returns for the CFP model and the GE stock prices, showing that in both cases a persistent autocorrelation exists and that it decays quite slowly, staying above zero for any lag's size, even if with different positive values (the GE values oscillate around an average value greater than that of CFP) . 

\subsubsection{Other Features}

{\it Stationarity}

\noindent It is worth to notice that our model generates simulated returns that show compatibility also with respect to another recurrent characteristic: stationarity on large time windows and non-stationarity on small intervals, \cite{mantegna1999introduction}. In order to check this feature we considered some returns series generated by the model, and tested the unit root hypothesis on series of different lengths, from very long to very short. More precisely, we tested the existence of a unit root in a number of series generated by the CFP model, with the following length:  $15000$ values, $1500$ values, and, finally, $150$ values. All of the series have been obtained by splitting the first one in smaller parts: i.e. ten series with $1500$ values have been obtained by splitting the series with $150000$ values in ten, and one hundred series with $150$ values have been obtained by splitting each of the $1500$ values long series in ten. In such a way we could test series of different length, without changing any structural feature of the dataset; none of the series is overlapping. Selected results of the augmented-Dickey-Fuller (ADF) test confirm that, both at $1\%$ and $5\%$ significance levels, the hypothesis of stationarity can be rejected for small time intervals, whereas a robust indication of stationarity exists for longer series. None of the $150$-values long series exhibited stationarity, and none of the others exhibited non-stationarity.

\bigskip

\begin{figure}
\centering 
\includegraphics[width=3.0in,angle=0]{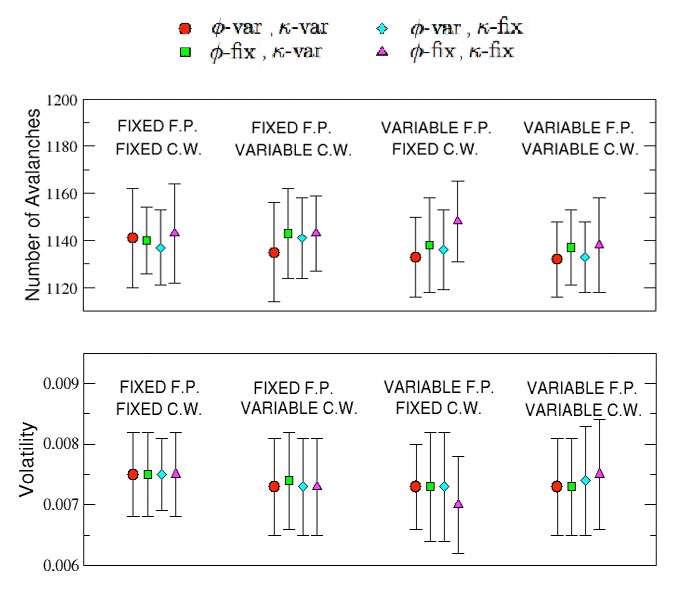} 
\caption{(Color online) Stability of the CFP model. The average number of avalanches obtained in simulations (top panel) are reported together with the average volatility values of the global price series (bottom panel) for various combinations of parameters. Vertical bars measure the correspondent standard deviations, over $50$ events. In all cases, variability of values observed lays within the standard deviation.}
\end{figure}

\noindent {\it Robustness}

\noindent The model appears also interestingly robust with respect to its parameters variability. As described in previous sections, the relevant parameters of the model are: (i) the fundamental price (either fixed or variable); (ii) the length of the chartists window (either fixed or variable); (iii) the values of both the ``sensitivity multipliers" for fundamentalists and chartists, namely $\phi$ and $\kappa$ (both of them can be either fixed or variable). We monitored in particular two indicators, namely the volatility of the returns series and the number of relevant avalanches (i.e. avalanches that involve at least the $0,2\%$ of traders), and run 50 simulations with different initial conditions for any parameters combination, each simulation consisting of 10000 tick prices. Fig.6 shows the effect of the parameters setting on the volatility of the global price series (bottom panel) and on the number of significative extreme events occurred during the simulations (top panel). Results suggest a strong stability of the artificial market operativeness, which can be easily read by considering that fluctuations of the indices around their average values are always smaller than their correspondent standard deviations. 

\begin{figure}[t]
\centering 
\includegraphics[width=3.5in,angle=0]{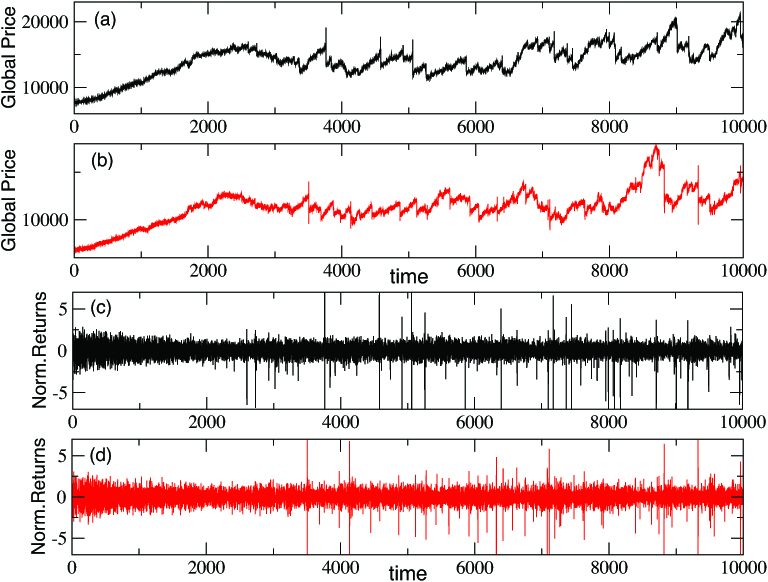} 
\caption{(Color online) Global prices and returns time series versus time, as in Fig.3, but for two decreasing percentages of chartists, i.e. $50\%$ , panels (a) and (c), and $25\%$, panels (b) and (d). The reduction of chartists does not imply significative changes in both the series with respect to the situation with $75\%$ of chartists shown in Fig.3.}
\end{figure}

\bigskip

\noindent {\it Population composition and network size}

\noindent Finally, let us briefly check how the behavior of the global price time series, described in subsection $2.2$ for a community of $N=1600$  traders with $75\%$ of chartists, are affected by a change in the relative composition of fundamentalists and chartists. In Fig.7 we show the same plots as in Fig.3, but for communities with, respectively, $50\%$ and $25\%$ of chartists: it is evident that both the global price time series and the returns series hold their features, even in case of a strong decrease of the percentage of chartists. 

This can be seen also by looking at Fig.8(a), where the pdf of normalized returns, calculated over series of $30000$ tick prices (in order to have a better statistics), is reported for different population compositions. The only noticeable difference with respect to the case shown in Fig.4(a), here reported again for comparison (as circles), is the central part of the distribution, which assumes a more rounded shape. Thus, the population composition does not affect \textit{per se} the fat tails characteristic of the returns distribution, which seem to be rather robust with respect to a variation in the percentage of the number of chartists. We checked that both fat tails and asymmetry in the returns distributions are robust also with respect to a variation in the size of the SW network, as shown in Fig.8(b), where the different pdfs refer to $N=2500$ and $N=3600$ (the case $N=1600$ is also reported for comparison). 

However, as we will show in the next section, things change a lot if we take into account our third category of traders, i.e. those who invest in a random way.

\begin{figure}
\centering 
\includegraphics[width=3.0in,angle=0]{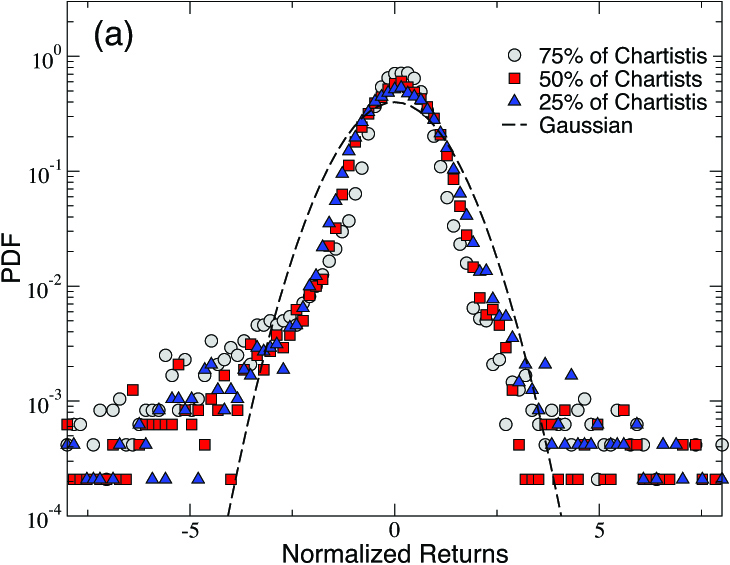} 
\includegraphics[width=3.0in,angle=0]{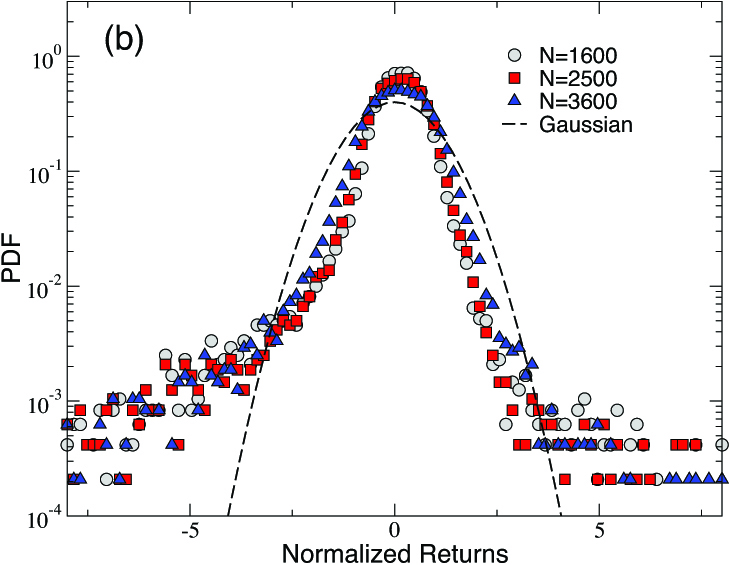} 
\caption{(Color online) Pdf of normalized returns for different agents compositions (a) and different population sizes (b). Fat tails and asymmetry in the returns distribution are not significantly affected neither by the percentage of chartists nor by the network size. See text.}
\end{figure}

\bigskip

\section{Random Traders and Policy Suggestions}

In previous studies \cite{noiPRE}, within a  simpler artificial financial market without feedback effects, it has been shown how the introduction of a small percentage of random traders affects the distribution of financial formationavalanches: more precisely, we showed that their presence in the trading community, even in a minimal proportion, is able to diminish the size of avalanches changing the power law behavior of pdfs into and an exponential one. Therefore we want to explore, now, the extent to which an analogous effect could be observed also in the returns distribution of the price series endogenously generated by traders in the context of CFP model. We will show that such an expectation is actually verified. 

\begin{figure}[t]
\centering 
\includegraphics[width=3.5in,angle=0]{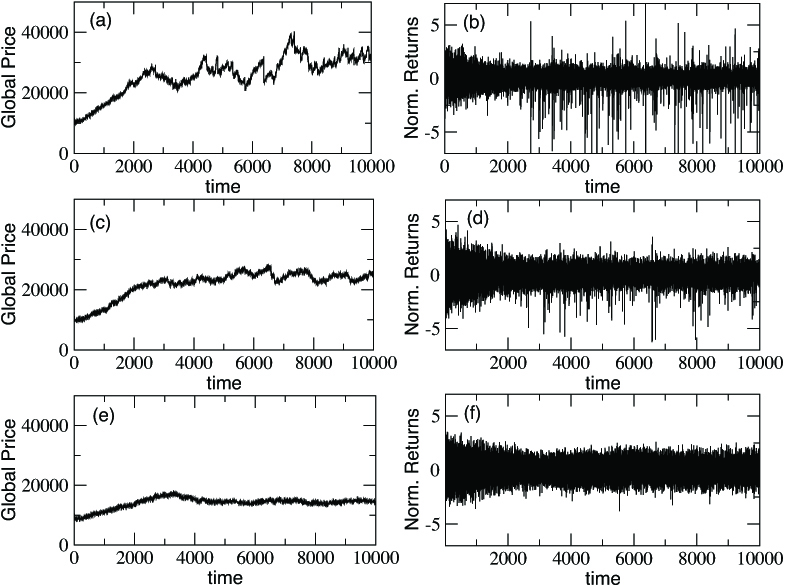} 
\caption{Global prices and normalized returns time series for three increasing percentages of random traders: $1\%$, panels (a),(b); $5\%$, panels (c),(d); $15\%$, panels (e),(f). The presence of random traders affects the structure and the properties of both time series reducing the occurrence of wild fluctuations. See text.}
\end{figure}

In order to do this, we introduce different percentages of random traders in the community, while the remaining part of non random agents is always divided in $75\%$ chartists and $25\%$ fundamentalists. 
We report in Fig.9 and Fig.10 the effect of the presence of random traders on the global price series, on the returns series and on the returns probability distribution. More precisely, in Fig.9 we show how the price time series is modified by increasing the percentage of random traders and how both the frequency and the magnitude of bursts in the returns series are strongly dampened already for small percentages (between $1\%$ and $5\%$). Such an effect is confirmed in Fig.10, where fat tails in the distribution of returns (calculated, again, over $30000$ tick prices) appear visibly reduced in presence of random traders. In particular, one observes a change in the shape from an asymmetric fat-tailed curve, still observed for $1\%$ of random traders, to a  less sharpened curve that appears already around $5\%$ and rapidly tends towards a standard Gaussian curve, finally obtained for percentages above $15\%$. A comparison of these results with the analogous reported in Fig.3 (panels b-d) and Fig.4$(a)$, where no random traders were present, clearly corroborates what we expected: the introduction of an increasing percentage of traders which fix their price at random, as described by Eq.3, reduces extreme events, in size and frequency, and induces a change in the shape of the returns distribution. It is quite remarkable  that a significative effect is obtained even with a sensible low percentage. Therefore, with respect to previous findings\cite{noiPRE}, we have obtained a further confirmation of the beneficial effects of a random trading behavior, which hold also in the more realistic context of the CFP model, where agents have a realistic feedback on the market. 

\begin{figure}[t]
\centering 
\includegraphics[width=3.0in,angle=0]{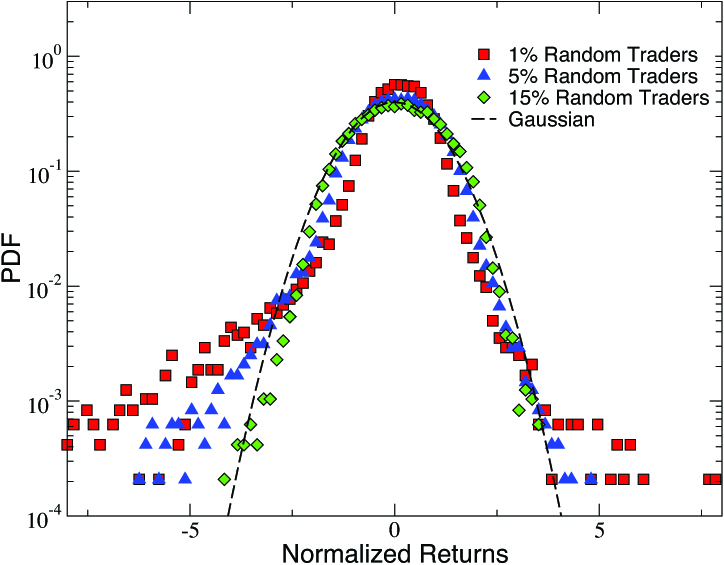} 
\caption{(Color online) Pdf of normalized returns for the same three percentages of random traders of the previous figure. The presence  of a small percentage of this kind of traders significantly reduces the fat tails of the returns distributions. Moreover, one observes also  an interesting reduction  in the asymmetry of the distribution for a population composition with $15\%$ of random traders, a value above which the pdfs become Gaussians, see text.}
\end{figure}

The results discussed above indicate  some counterintuitive policy suggestions. Although it may appear quite provocative  to state that random traders may dampen both size and frequency of avalanches, these results are quite robust and have been confirmed in several situations. 
On the other hand, it is a medium-level target of financial policy to stabilize the market and reduce wild fluctuations.  The presence of a small number of random traders seem to be able to produce a similar  effect. In fact we have seen that they are able to produce financial returns which tends to be normally distributed. In this way the probability to predict price dynamics increases or, at least, the error of forecasts might be reduced. But  what is it \emph{actually} the meaning of traders who acts in a random way?  One of the most straightforward explanation is that, in real markets, information flows determines decisions of investors. Then, the presence of a few  random players simply limits the diffusion of information, providing boundaries to the spreading mechanism. 

The paradoxical extreme of contemporary financial markets is that there exists an excess of information that is the cause of turbulent and unpredictable dynamics. What a single trader does not know, is exactly the content of the informative signal that she will try to obtain. This induces a continue search for credible signals and, once a \emph{presumably} credible signal is found,  it starts  the spreading of the contagion through avalanches of any size. The presence  of random traders may provide  two interesting effects: from an individual point of view, as already suggested in \cite{noiPRE,noiPLOSONE}, a random approach to financial speculative investments may reveal to be more convenient and less risky; on the other hand  from a collective perspective, the presence of traders who do not carry any signal to follow can reduce the informational cascade that generates contagion. A counterintuitive  policy suggestion can then be derived: the amount of information that traders \emph{believe} to retrieve from the market must be limited. The limitation will act exactly as the effect of our random investors in the CFP model: it will reduce the fragility of the market, the continuous search for signals, the unstable and dangerous reactivity of investors, ready to imitate any possibly credible behavior. Markets dynamics will result to be closer to equilibrium without dangerous  extreme fluctuations and easier to predict.

\section{Conclusions}
We have presented a new model, the CFP model, for an artificial financial market with heterogeneous agents. By means of a SOC-like herding behavior of agents, the CFP model is able to provide realistic time price series that reproduce well known stylized facts and that compare well with real time series. We have also investigated how wild price fluctuations can be damped by introducing a new category of agents who trade in a random way. Promising results  have been obtained in this direction, confirming  some previous findings related to the beneficial role of random strategies. Even the introduction of a small percentage of these random trading agents is able to diminish wild price fluctuations. In this respect the reduction of information seems to be a  convenient although counterintuitive policy suggestion for market stabilization which deserves a more detailed investigation.  

\section{Acknowledgements}
We thank Ugur Tirnakli and Constantino Tsallis for useful discussions. This work was partially supported by the INFN Project "PIECES" and by the FIR Research Project 2014 N.ABDD94 of Catania University.


\begin{references}
\bibitem{generalelectric}
This time series has been provided by Thomson Reuters.
\end{references}
\end{document}